\documentclass[runningheads,a4paper]{llncs}
\usepackage{color}
\usepackage{amssymb}
\usepackage{graphicx}
\usepackage{framed}
\usepackage{url}
\usepackage{etoolbox}

\BeforeBeginEnvironment{figure}{\vskip-4ex}
\AfterEndEnvironment{figure}{\vskip-2ex}

\urldef{\mailsa}\path|firstname.lastname@gesis.org|
\newcommand{\keywords}[1]{\par\addvspace\baselineskip
\noindent\keywordname\enspace\ignorespaces#1}

\begin{document}

\mainmatter  

\title{Evaluating Co-Authorship Networks in Author Name Disambiguation for Common Names}


%
%
\author{Fakhri Momeni%
\and Philipp Mayr}
\titlerunning{Author Name Disambiguation for Common Names}

\author{Fakhri Momeni and Philipp Mayr}
\institute{GESIS Leibniz-Institute for the Social Sciences, Cologne, Germany \\
	\email{firstname.lastname@gesis.org} }

%
%

\maketitle

\begin{abstract}
With the increasing size of digital libraries it has become a challenge to identify author names correctly. The situation becomes more critical when different persons share the same name (homonym problem) or when the names of authors are presented in several different ways (synonym problem). This paper focuses on homonym names in the computer science bibliography DBLP. The goal of this study is to evaluate a method which uses co-authorship networks and analyze the effect of common names on it.
For this purpose we clustered the publications of authors with the same name and measured the effectiveness of the method against a gold standard of manually assigned DBLP records. The results show that despite the good performance of implemented method for most names, we should optimize for common names. Hence community detection was employed to optimize the method. Results prove that the applied method improves the performance for these names.

\end{abstract}

\section{Introduction} 
In scholarly digital libraries (DLs) authors are recognized via their publications. 
It is important for users to know about the author of a particular publication to access possible other publications of this author.
For this purpose DLs provide search services using the publication information in their databases. 
However, when several authors share the same name or authors provide their works with different names DLs need more analysis on author's oeuvres.  
Many different approaches have been proposed in the field of author name disambiguation. 
Manual author identification in large DLs is very costly. The consequence is that large sets of ambiguous author names need to be analyzed automatically. 
In addition the demographic characteristics such as name origin and frequency of names used for authors influence the identification of authors. 
Therefore, all constrains of the underlying data should be considered to choose the appropriate method for author name disambiguation.

Author assignment method and author grouping method \cite{Anderson:methods} are the two main methods for author name disambiguation. Author assignment method constructs a model that represents the author and assigns proper publications to the model. It requires former knowledge about the authors. Nguyen and Cao  \cite{cao:assignmentMethod} used this method and proposed to link the author names to the matching entities in Wikipedia.
The author grouping method clusters the publications on the basis of their properties (co-authors, publication year, keywords, etc.) to assign a group of publications to a certain author.
Following this framework, Caron and van Eck \cite{Caron:rulebase} applied rule-based scoring to clustered publications. In their approach the authors suppose that there is enough information about authors and their documents. Also, Gurney et al. \cite{Gurney:communitydetection} clustered publications with employing different data fields and integrated a community detection method. 
Some authors \cite{Felipe:socialnetEvaluate},\cite{Shin:socialnet},\cite{Peng:socialnet} used social networks (mainly co-authorship networks) to cluster publications. Levin and Heuser \cite{Felipe:socialnetEvaluate} introduced a set of matching functions based on the social network of authors and measured the strength of connections between the authors. Shin et al. \cite{Shin:socialnet} extracted the abstract and author's affiliation from the paper and considered the relation between authors to find similarities between publications. Wang et al. \cite{Peng:socialnet} proposed a unified semi-supervised framework to handle the synonym and homonym problem of author names.

In this paper we used an author grouping method (compare \cite{Anderson:methods}) to cluster the publications of a set of random authors with the same name in the DBLP database. Considering the lack of rich bibliographic information in DBLP records, we applied co-authorship network analysis introduced by Levin and Heuser \cite{Felipe:socialnetEvaluate} to detect similarities between publications in order to investigate, how the amount of homonym names affects the disambiguation results. In the end, we employed a community detection algorithm (Louvain method) to reduce the effect of common names in our evaluation.

\section{Disambiguation Approach}\label{sec:disApp}     
We use the author grouping method in order to assign all publications of each person to a certain group. For this purpose all publications belonging to the same ambiguous author name are categorized into one block. In a next step we compare any pair of publications in each block with each other to find a similarity between them. If we have \textit{n} blocks and $m_i$ publications in a block \textit{i}, the number of comparisons for all blocks is:
\begin{equation}\sum\limits_{i=1}^n \frac{m_i(m_i-1)}{2}\end{equation}
The result of each comparison is true or false. The \textit{true} result means that two publications belong to one person and the same cluster. If one of them was compared with another one before and assigned to a cluster, the other one is added to that cluster too. If both of them were compared before and belong to different clusters, two clusters are rebuilt to one cluster. Otherwise a new cluster will be created and two publications are put in new cluster. In the next section we describe how to define the similarity indicator to build the clusters.  

\begin{figure}[h]                                                     
    	\includegraphics[scale=0.25]{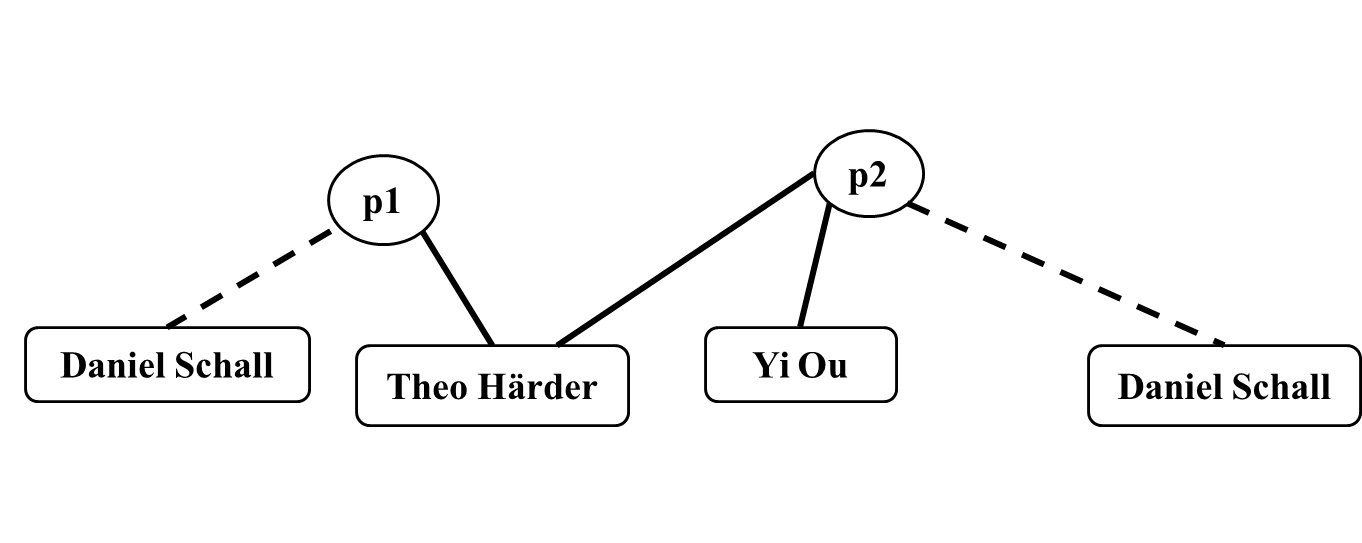}%
    	\hspace{4mm}%
    	\includegraphics[scale=0.25]{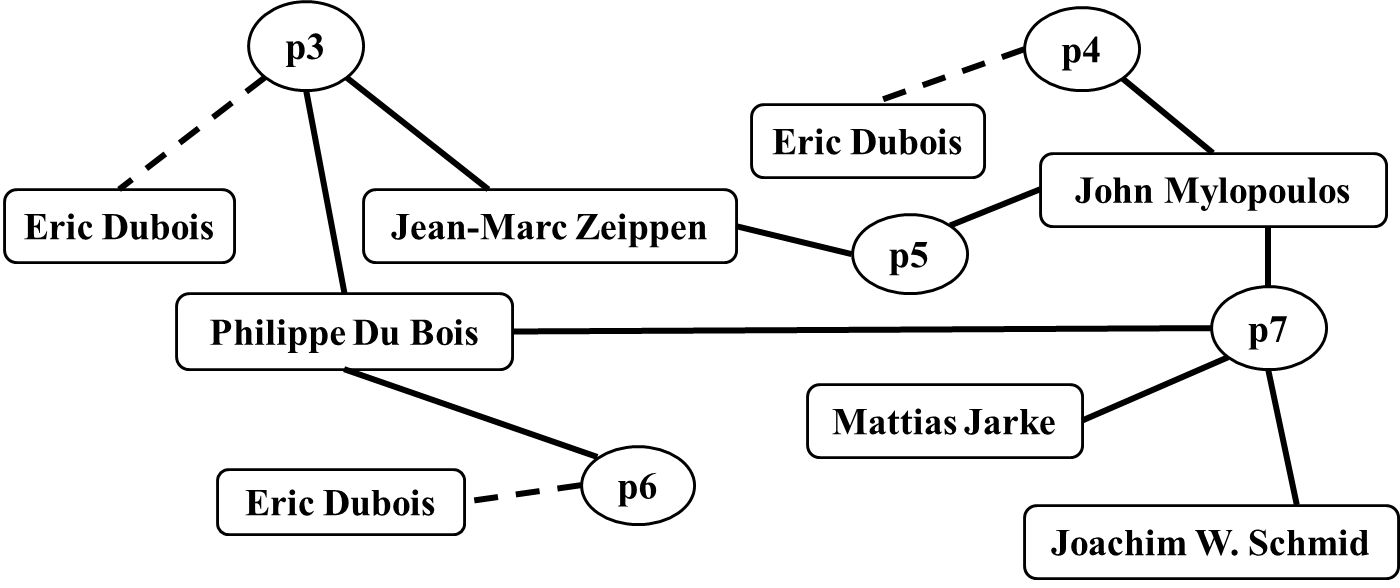}%
    	\caption{An example of a co-authorship network}
    	\label{fig:dblpnet}
    \end{figure}

The bibliographic information that we can obtain from publications in DBLP is limited mainly to author names (the names of all co-author names are listed), title and publication venue. We chose the co-author names as our similarity indicator. Therefore we built a network of authors and documents. 
Figure \ref{fig:dblpnet} shows an example of the network. The continuous lines show the links between publications and authors in the network.
As it was mentioned before each pair of documents within every block has to be compared.
To compare the publications the relations in the network are analyzed. 
If there is a path between two publications, their distance is defined as the length of the shortest path between them, otherwise it would be infinite. The length of the shortest path is equal to the number of nodes between two nodes. For example the distance between publication p1 and p2 in Figure \ref{fig:dblpnet} is 1; the distance between p3 and p4 is 3.
The less distance between two publications means that more likely these publications were written by one person. So, the distance between two publications is assumed as the similarity measure. Different thresholds can be considered for the distance. For example, in Figure \ref{fig:dblpnet}, with the threshold = 1, p1 and p2 are two publications of one person with the name 'Daniel Schall', because they share the same co-author. Accepting the threshold = 3, p3 and p4 belong to same author with the name 'Eric Dubois'.
In Section \ref{sec:evaluation} we see the effect of selecting different thresholds on the evaluation results.   
  
\section{Evaluation} \label{sec:evaluation} 
\textbf {Gold Standard}: In order to evaluate the output of the author disambiguation approach we need a gold standard of disambiguated author names. 
Many homonym author names in DBLP are disambiguated manually by the DBLP team and are identifiable with an id. For example, 'Wei Li' belongs to 59 different persons: 'Wei Li 0001', 'Wei Li 0002', etc. Thus, the set of publications for each person is recognizable. To build the gold standard \cite{Momeni2016} we selected these identified author names and compiled all their publications into one set. In our gold standard we provide a list of publications which have at least one disambiguated author name.
Asian names, especially Chinese names are the most common names in DBLP and result in many homonym author names. These names are the most problematic names in author disambiguation  
 and should be analyzed in particular. 
In total 1,578,316 unique author names exist in DBLP. There are 5,408 authors who have an identification number (we mention them as disambiguated authors). These 5,408 authors and their publications form the gold standard. We got these numbers from DBLP, downloaded May/01 2015 from http://dblp.uni-trier.de/xml/.
To measure the performance of our method 1,000 disambiguated author names have been randomly selected from the gold standard. In total we have 2,844 different authors and 32,273 publications in our random sample. 
In the next section we evaluate the performance of our method against the gold standard.

\textbf{Evaluation Metrics}:
Bcubed metrics \cite{Enrique:Bcubed} are used to evaluate the quality of the algorithm. These clustering metric satisfy constraints on evaluation the clustering tasks \cite{Enrique:Bcubed} such as cluster homogeneity and cluster completeness. 
Therefore we applied them to evaluate our method. For this purpose Bcubed precision and recall are computed for each publication.
\begin{figure}
	\centering
	\includegraphics[scale=0.25]{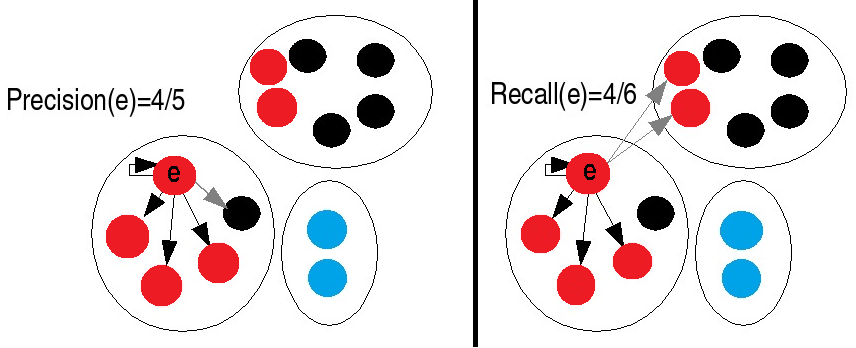}
    \caption{Example BCubed precision and recall adapted from \cite{Enrique:Bcubed}}
	\label{fig:Bcubed}
\end{figure}
Figure \ref{fig:Bcubed} shows an example how the precision and recall of one publication of an author are computed by BCubed metrics. In this Figure assume the circles in red, black and blue as the publications belonging to three different authors and our algorithm categorized them to three groups. The publication precision measures how many publications in its group belong to its author. The publication recall measures how many publications from its author appear in its group. 
Bcubed F as the combination of Bcubed precision and recall is computed as follows:
\begin{equation}\frac{1}{\alpha(\frac{1}{P})+(1-\alpha)(\frac{1}{R})}\end{equation}
being P and R Bcubed precision and recall and being $\alpha$ and (1- $\alpha$) the relative weight of each metric (We assumed $\alpha$ = 0.5).
Bcubed precision, recall and F-measure were computed for every publication in any block. Then we consider their average as the Bcubed precision, recall and F of the block.
	
	\section{Results and Discussion}\label{sec:result}
We clustered the publications with regard to the distance between them. For choosing the threshold we have checked the distances larger than 3, which results in a very low precision. Then we chose the threshold equal to 1 and 3. For the distance less than threshold (1 or 3), we assign two publications in the same cluster.	
The results of the evaluations for two thresholds are demonstrated in Table \ref{tab:evaluation}. 
		\begin{table}\vspace{-1em}
		\centering
		\caption{Mean values of BCubed metrics for 1,000 blocks } \label{tab:evaluation}
		\begin{tabular}{| p{2.5cm} | p{3cm} | p{3cm} | p{2.5cm} |}
			\hline
			 &\centering BCubed precision&\centering BCubed recall& \centering BCubed F \tabularnewline \hline
						\begin{tabular}{l}
						Threshold=1
						\end{tabular}
			 & \centering0.98 & \centering0.74 & \centering 0.79 \tabularnewline 
			  \hline
			  			 \begin{tabular}{l}
			  			 	Threshold=3
			  			 \end{tabular} & \centering0.94 & \centering0.81 & \centering 0.82 \tabularnewline  
			\hline
		\end{tabular}
	\end{table}
The results in Table \ref{tab:evaluation} indicate that our co-author networks method performs well on the dataset and it can be utilized as author identification approach. No effort was made to define and compare against an external baseline. 
Comparing the results for two thresholds (1 and 3) we can conclude that using threshold=3 provides us with the better balance between precision and recall and a higher F (slightly better BCubed recall of 0.81 and F of 0.82).
We can shows that with the increasing number of publications in the blocks, the efficiency of our algorithm decreases, especially for threshold=3.  
We can conclude that although using threshold=3 results the better performance generally, it is less efficient than using threshold=1 for common names.    
The reason is that common names enhance the probability of being authors with the same name in the same area of research activity and increase the likelihood of detecting the shared co-author for different researchers with the same name. 
Furthermore, it is more likely that these authors have co-authors with similar common names.
This results in a higher probability of ambiguous co-authors and wrong connections between publications.
Therefore, we should be more cautious to use the co-author of co-author as the similarity measure for these cases and will verify them more deeply. 
To remove the wrong connections that link two groups of publications from different authors community detection is a good solution. 
\emph{Community detection} aims at grouping nodes in accordance with the relationships among them to form strongly linked subgraphs from the entire graph. 
Hence, we applied a community detection algorithm to optimize the results (threshold=3) for the common names. We chose a subset of the author's names which have more than 200 publications (totally 28 names) in our DBLP dataset.
To detect communities in the network we utilized the Louvain method in Pajek. This method maximizes the modularity of network. Single refinement is selected and the resolution parameter was set to 1. 
Because the less distance between publications increases the probability of being the same author, we gave the weight to connections. For the distances equal to 1 and 2 have weights with values 2 and 1 respectively. 
Table \ref{tab:communityCommonname} shows that community detection improved the results for the most repeated names in our sample.
	\begin{table}\vspace{-1em}
		\centering
		\caption{BCubed metrics for author names with more than 200 publications, thr.=3} \label{tab:optimize}
		\label{tab:communityCommonname}
		\begin{tabular}{| l | p{3cm} | p{2.5cm} | p{2.5cm} |}
			
			\hline
			& \centering BCubed precision & \centering BCubed recall & \centering BCubed F \tabularnewline  \hline

						\begin{tabular}{l}
							Before  optimization
						\end{tabular} & \centering0.46 & \centering0.87 & \centering0.45 \tabularnewline  
			\hline
		\begin{tabular}{l}
			After optimization
		\end{tabular}	& \centering0.79 & \centering0.61 & \centering0.58\tabularnewline      
			\hline
		\end{tabular}
	 \end{table} 
\section{Conclusions and Future Work}\label{sec:conclusion}
In this paper we implemented a method to identify authors with the same name based on co-authorship networks in DBLP. The results showed that although co-author networks have a substantial impact on author name disambiguation, but common names decrease the performance of our method and should be optimized in an extra step. For this reason, we implemented the community detection method which showed an improvement for highly frequent common names. 
Our approach can be applied to disambiguate author names in DBLP. In this way we create the network and link the publications automatically, then apply the community detection to find the suspicious connections and check them manually if they are a wrong connection. In this case, they will be removed from the network and increase the performance of algorithm. So, the speed of automatic disambiguating and the accuracy of manual checking can be combined. Our approach improves the disambiguation of common names, but this is not sufficient. To get better results we need to optimize the parameters such as resolution in the community detection method for different numbers of publications per name. We could also investigate the effect of changing the weights of links between publications depending on their distances.
Because this method is based on co-author network, it is limited to multi-author papers. Therefore a multi-aspect indicator is required for single author papers. 
We can use the titles of publications to extract keywords and add this information to calculate similarity measures.
\section{Acknowledgment}\label{sec:ACKNOWLEDGMENTS}
This work was funded by BMBF (Federal Ministry of Education and Research, Germany) under grant number 01PQ13001, the  Competence Centre for Bibliometrics.
We thank our colleagues at DBLP who helped with generating the testbed \cite{Momeni2016}.


\bibliographystyle{splncs03}
\bibliography{tpdl}

\begin{thebibliography}{1}
\providecommand{\url}[1]{\texttt{#1}}
\providecommand{\urlprefix}{URL }

\bibitem{Enrique:Bcubed}
Amig{\'{o}}, E., Gonzalo, J., Artiles, J., Verdejo, F.: A comparison of
  extrinsic clustering evaluation metrics based on formal constraints. Inf.
  Retr.  12(4),  461--486 (2009)

\bibitem{Caron:rulebase}
Caron, E., van Eck, N.J.: Large scale author name disambiguation using
  rule-based scoring and clustering  (2014)

\bibitem{Anderson:methods}
Ferreira, A.A., Gon{\c{c}}alves, M.A., Laender, A.H.F.: A brief survey of
  automatic methods for author name disambiguation. {SIGMOD} Record  41(2),
  15--26 (2012)

\bibitem{Gurney:communitydetection}
Gurney, T., Horlings, E., den Besselaar, P.V.: Author disambiguation using
  multi-aspect similarity indicators. Scientometrics  91(2),  435--449 (2012)

\bibitem{Felipe:socialnetEvaluate}
Levin, F.H., Heuser, C.A.: Evaluating the use of social networks in author name
  disambiguation in digital libraries. {JIDM}  1(2),  183--198 (June 2010)

\bibitem{Momeni2016}
Momeni, F., Mayr, P.: {An Open Testbed for Author Name Disambiguation
  Evaluation} (2016), \url{http://dx.doi.org/10.7802/1234}

\bibitem{cao:assignmentMethod}
Nguyen, H.T., Cao, T.H.: Named entity disambiguation: {A} hybrid statistical
  and rule-based incremental approach. In: {ASWC} 2008

\bibitem{Shin:socialnet}
Shin, D., Kim, T., Jung, H., Choi, J.: Automatic method for author name
  disambiguation using social networks (2010)

\bibitem{Peng:socialnet}
Wang, P., Zhao, J., Huang, K., Xu, B.: A unified semi-supervised framework for
  author disambiguation in academic social network. In: {DEXA} 2014

\end{thebibliography}

\end{document}